\documentclass[pra,letterpaper,twocolumn,showpacs,superscriptaddress,floatfix,longbibliography]{revtex4-1}
\usepackage{graphicx,psfrag,amsmath,amssymb,amsfonts,bbm,latexsym,color,dcolumn,bm,mathbbol,mathrsfs}
\allowdisplaybreaks 

\newcommand{\bra}[1]{\langle #1 |} \newcommand{\ket}[1]{| #1 \rangle}
\newcommand{\I}{\mathrm{i}} \newcommand{\D}{\mathrm{d}}

\def\RE{\mathop{\rm Re}} %
 \definecolor{myred}{RGB}{168,5,14}
\definecolor{myblue}{RGB}{13,13,255}
\definecolor{mygreen}{RGB}{20,150,20}

\begin{document}

\title{ Thermalization of noninteracting quantum systems coupled to
  blackbody radiation:\\ A Lindblad-based analysis }

\author{Massimo Ostilli} \affiliation{Departamento de F\'isica
  T\'eorica e Experimental, Universidade Federal do Rio Grande do
  Norte, 59078-970 Natal-RN, Brazil} \affiliation{Departamento de F\'isica,
  Universidade Federal de Santa Catarina, 88040-900 Florianop\'olis,
  SC, Brazil} \author{Carlo Presilla} \affiliation{Dipartimento di
  Fisica, Sapienza Universit\`a di Roma, Piazzale A. Moro 2, 00185
  Roma, Italy} \affiliation{Istituto Nazionale di Fisica Nucleare,
  Sezione di Roma 1, 00185 Roma, Italy}

\date{\today}

\begin{abstract}
  We study the thermalization of an ensemble of $N$ elementary,
  arbitrarily-complex, quantum systems, mutually noninteracting but
  coupled as electric or magnetic dipoles to a blackbody radiation.
  The elementary systems can be all the same or belong to different
  species, distinguishable or indistinguishable, located at fixed
  positions or having translational degrees of freedom.  Even if the
  energy spectra of the constituent systems are nondegenerate, as we
  suppose, the ensemble unavoidably presents degeneracies of the
  energy levels and/or of the energy gaps.  We show that, due to these
  degeneracies, a thermalization analysis performed by the popular
  quantum optical master equation reveals a number of serious
  pathologies, possibly including a lack of ergodicity. On the other
  hand, a consistent thermalization scenario is obtained by
  introducing a Lindblad-based approach, in which the Lindblad
  operators, instead of being derived from a microscopic calculation,
  are established as the elements of an operatorial basis with squared
  amplitudes fixed by imposing a detailed balance condition and
  requiring their correspondence with the dipole transition rates
  evaluated under the first-order perturbation theory.  Due to the
  above-mentioned degeneracies, this procedure suffers a basis
  arbitrariness which, however, can be removed by exploiting the fact
  that the thermalization of an ensemble of noninteracting systems
  cannot depend on the ensemble size. As a result, we provide a
  clear-cut partitioning of the thermalization time into dissipation
  and decoherence times, for which we derive formulas giving the
  dependence on the energy levels of the elementary systems, the size
  $N$ of the ensemble, and the temperature of the blackbody radiation.
\end{abstract}

\maketitle
\section{INTRODUCTION}
The study of open quantum systems \cite{Petruccione,Schaller} is
crucial for a modern understanding of the foundations of quantum
mechanics, and finds numerous applications in fields such as solid
state physics, quantum optics, and quantum computation, where
questions about decoherence and dissipation are of theoretical and
practical importance. In particular, the call for nonunitary
evolutions designed to perform quantum operations, like purification
via, e.g., quantum adiabatic algorithms \cite{Farhi2001}, or cooling
via, e.g., thermalization~\cite{Scardicchio}, raises issues on the
possibility of preparing a large system of $N$ qubits in its ground
state \cite{Viola}.

In the above and other situations, one deals with the problem of
evaluating the thermalization time $\tau$ of a quantum many-body
system in contact with a thermal reservoir.  To be more precise, let
$\bm{H}$ be the Hamiltonian of the isolated many-body system and
$\bm{\rho}(t)$ its reduced density matrix operator at time $t$, when
interacting with a thermal reservoir at temperature $T$.  Ordinary
quantum statistical mechanics postulates that, upon reaching thermal
equilibrium, the system acquires two main features: (a) it becomes an
incoherent mixture of its eigenstates, i.e., for $t\to\infty$,
$\bm{\rho}(t)$ becomes diagonal in any eigenbasis of $\bm{H}$; and (b)
the Gibbs distribution is attained, i.e., $\bm{\rho}(t)\to
\bm{\rho}^{(\mathrm{eq})} = e^{-\beta \bm{H}}/\mathrm{tr}(e^{-\beta
  \bm{H}})$, where $\beta=1/(k_BT)$, $k_B$ being the Boltzmann
constant.  Two natural questions then arise: Given $T$, and given the
initial density matrix operator $\bm{\rho}(0)$, how can $\bm{\rho}(t)$
evolve toward the Gibbs distribution?  and What are the typical
relaxation times $\tau^{(Q)}$ and $\tau^{(P)}$ required for (a) and
(b) to be established, respectively?

A milestone approach to open quantum systems is provided by the
Lindblad equation~\cite{Lindblad,GKS}.  A crucial point in the
Lindblad equation is the choice of the involved Lindblad operators. As
long as these operators remain unspecified, we refer to the Lindblad
class.  More precisely, we have a Lindblad class for any given system
characterized by its Hamiltonian $\bm{H}$.  It is known that there is
no Lindblad equation able to reproduce the exact evolution of
$\bm{\rho}(t)$, for the dynamics of an open system is never really
linear in $\bm{\rho}(t)$~\cite{Grabert1982,Grabert1983}.
Nevertheless, a Lindblad equation can provide an effective description
in terms of a coarse grained dynamics, if the correlation time of the
environment is much shorter than the correlation time of the isolated
system~\footnote{It has been demonstrated that this condition is
  certainly met in the limit of weak coupling between system and
  environment~\cite{Davies1974}, or when the system is coupled to an
  infinite free boson bath with Gaussian correlation functions having a
  vanishing decay time~\cite{GK1976}.  It follows that for the
  specific environment we will consider here, namely, a blackbody
  radiation, the use of a Lindblad equation seems well justified due
  to both the weakness of the electromagnetic coupling constant and
  the almost complete incoherence of the thermally equilibrated
  electromagnetic modes~\cite{MethaWolf1964I}.}.  By varying the
choice of the Lindblad operators in the Lindblad class, one can span a
large variety of different Lindblad equations, among which an optimal
one can be found in terms of closeness to the real $\bm{\rho}(t)$ of
the given system and environment. We call this method of
determining the optimal Lindblad equation once the Lindblad class has
been postulated, the Lindblad-based approach (LBA).  The optimization
procedure could require nontrivial conditions to be imposed.

Parallel to the LBA, there exists a more physically motivated method
for proceeding, namely, a microscopic derivation.  Given the system,
the environment, and the system-environment Hamiltonians, by explicit
calculations one obtains an equation for $\bm{\rho}(t)$ that belongs
to the Lindblad class with the Lindblad operators fully determined.  A
celebrated example is the quantum optical master
equation~\cite{Petruccione,Schaller}, which describes the dipole
interaction of a system with a blackbody radiation.  However, in the
microscopic derivation a series of approximations is introduced in
order to reach an equation that belongs to the Lindblad class. Even if
these approximations are physically sound, their validity is, in
general, out of control.  It may happen that, even if the equation
resulting from the microscopic derivation belongs to the Lindblad
class, it does not coincide with the Lindblad equation found in the
LBA.

In this paper, we study the approach to equilibrium of an ensemble of
$N$ \textit{noninteracting} elementary quantum systems coupled as
electric or magnetic dipoles to a blackbody radiation.  As long as the
interaction with the radiation can be effectively reduced to dipolar
terms only, the elementary systems can be arbitrarily complex. They
can be all the same or belonging to different species, distinguishable
or indistinguishable, located at fixed positions or moving in a
lattice or in a box.  Examples range from a pure
gas~\cite{PhysRevLett.88.123001} or a
mixture~\cite{PhysRevA.91.022709} of chiral molecules at a low
density, to a chipset of superconducting flux qubits~\cite{Boixo2014}.
Ensembles of $N$ mutually noninteracting systems are nontrivial when
coupled to an environment.  As far as we know, there are no explicit
formulas describing their relaxation times to equilibrium when $N>1$.
Note that here the problem is not merely mathematical (there are
excellent methods to solve a given Lindblad equation when $\bm{H}$ is
quadratic and the Lindblad operators proportional to one-body jump
terms \cite{Prosen_2008,Prosen_2010}); we are faced with the more basic
problem of how to completely determine the Lindblad operators.  We
also compare the Lindblad equation obtained in our approach with the
quantum optical master equation and show that, whereas the former
provides a thermalization scenario consistent with phenomenology, the
latter has pathological behaviors.  This proposes the LBA also for
studying interacting systems where many-body jump operators are
involved~\cite{LMG_MOP,therm}.

\section{METHODS}
For a system with a nondegenerate energy spectrum, the LBA is
relatively simple.  Suppose that the system interacts with a blackbody
radiation. We define an optimal Lindblad equation by three steps: (a)
choose the Lindblad operators as the jump operators among all possible
energy levels of the isolated system, (b) impose a detailed balance
condition (existence of the Gibbs stationary state) among the squared
amplitudes of these jump operators, and (c) evaluate the latter
squared amplitudes as the rates for dipole transitions. The Lindblad
equation obtained in this way coincides with the quantum optical
master equation and respects the basic properties: (p1) the stationary
state (SS) is unique, and (p2) the SS coincides with the Gibbs state.

For ensembles of $N$ noninteracting systems, even if each system has
nondegenerate energy spectra, the procedure, steps (a)--(c), cannot be
repeated unaltered, for the spectrum of the ensemble is degenerate (if
two or more systems have equal spectra) and the eigenbasis of the
ensemble Hamiltonian is not unique.  When degeneracies are present,
there are infinite ways to define the Lindblad operators associating
them with the infinite different eigenbases.  Each choice can give
rise to different thermalization times.  However, we can remove this
arbitrariness by selecting a particular eigenbasis (and corresponding
Lindblad operators) and constructing with it a Lindblad equation as
above, if we impose a further obvious property: (p3) in an ensemble of
$N$ equal but distinguishable noninteracting systems, the
thermalization time does not depend on $N$.  The resulting Lindblad
equation, besides respecting (p1)-(p3), provides the following
scenario.  There exist two natural characteristic times, $\tau^{(P)}$,
representing the time by which the system loses or gains energy, and
$\tau^{(Q)}$, representing the time by which the system loses quantum
coherence. We have $\tau=\max\{\tau^{(P)},\tau^{(Q)}\}$, and for
$\beta$ finite and $N$ large enough we find: (p4)
$\tau^{(Q)}\leq\tau^{(P)}$, (p5) $\tau^{(Q)}=O(1/N)$, and (p6)
$\tau^{(P)}=O(1)$ for distinguishable systems (equal or not) and, at
low enough densities, for indistinguishable systems too [note that
(p5) and (p6) are consistent with (p3): $\tau=O(1)$].  We provide
explicit formulas for $\tau^{(P)}$ and $\tau^{(Q)}$ in the case of
distinguishable systems.  Depending on the degree of degeneracy, the
quantum optical master equation does not satisfy some or all of the
above properties.

\subsection{Lindblad-based approach} 
Consider a generic system described by a Hermitian Hamiltonian
operator $\bm{H}$ acting on a Hilbert space $\mathscr{H}$ of dimension
$M$.  We assume that the eigenproblem, $\bm{H} \ket{m} = E_m \ket{m}$,
has discrete, nondegenerate, eigenvalues. The set of the corresponding
eigenstates $\{\ket{m}\}$ is an orthonormal basis in $\mathscr{H}$.
We arrange the eigenvalues in ascending order $E_1 < E_2 < \dots <
E_M$.

When the system is coupled to a thermal reservoir, we assume that its
reduced density matrix operator $\bm{\rho}(t)$ is determined by a
generic Lindblad class equation~\cite{Lindblad,GKS}
\begin{align}
  \label{GEN_LIN}%
  \frac{\D \bm{\rho}}{\D t} = -\frac{\I}{\hbar}
  \left[\bm{H}',\bm{\rho}\right] +\sum_{\alpha} \left(
    \bm{L}_{\alpha}\bm{\rho}\bm{L}^{\dag}_{\alpha} -\frac{1}{2}\left\{
      \bm{L}^{\dag}_{\alpha}\bm{L}_{\alpha},\bm{\rho} \right\}
  \right).
\end{align}
In this equation, the coherent part of the evolution is represented by
the Hermitian operator $\bm{H}'$, which, in general, differs from the
isolated system Hamiltonian $\bm{H}$.  The Lindblad, or quantum jump,
operators $\bm{L}_\alpha$ are, for the moment, completely arbitrary.
Even their number is arbitrary but can always be reduced to $M^2-1$
\cite{Pearle}.

The set of $M^2$ dyadic operators $\{\ket{m}\bra{n}\}$ forms an
orthonormal basis in the space of the operators acting on
$\mathscr{H}$ equipped with the Hilbert-Schmdit scalar product
$\langle \bm{O}_1,\bm{O}_2\rangle=\mathrm{Tr}(\bm{O}^\dag_1\bm{O}_2)$.
In the case of dipolar interactions, and within the first-order
time-dependent perturbation theory, diagonal transitions
$|n\rangle\to|n\rangle$ are forbidden \cite{Davydov}, so that the
above set reduces to the set of the $M(M-1)$ non diagonal dyadic
operators.  We then identify this set as the set of the Lindblad
operators of Eq.~(\ref{GEN_LIN}), namely,
\begin{align}
  \label{Lpq}%
  \bm{L}_\alpha \to \bm{L}_{m,n} =\ell_{m,n}|m\rangle\langle n|,
  \qquad m\neq n.
\end{align}
The meaning of the $M^2$ complex coefficients $\ell_{m,n}$ is made
clear soon.  Note that, if the spectrum of $\bm{H}$ has degeneracies,
the eigenbasis $\{\ket{m}\}$ is not unique and neither are the jump
operators.  In this case, one should modify definition~(\ref{Lpq})
in such a way that the resulting theory is invariant under a change of
the basis $\{\ket{m}\}$.  We will discuss this point elsewhere.

Let us denote $\bra{m}\bm{\rho}\ket{n}=\rho_{m,n}$ and
$\bra{m}\bm{H}'\ket{n}=H'_{m,n}$.  On imposing that
$\bm{\rho}^{(\mathrm{eq})}$ is a stationary solution of
Eq.~(\ref{GEN_LIN}) with jump operators~(\ref{Lpq}), namely, that
property (p2) is satisfied, we get $H'_{m,n} = E'_{m} \delta_{m,n}$
and
\begin{align}
  \label{ellemn}%
  |\ell_{m,n}|^2 = C_{m,n} e^{-\frac{\beta}{2} (E_{m}-E_{n })},
\end{align}
with $C_{m,n}=C_{n,m}\geq 0$; see Appendix~\ref{LBA-appendix} for
details.  It follows that the $M$ diagonal elements $\rho_{m,m}$ have
an evolution decoupled from that of the $M(M-1)$ off-diagonal terms
$\rho_{m,n}$, $m\neq n$, which, in turn, are decoupled from each
other~\cite{AlickiLendi}.

On defining $\bm{p}$ as the $M$-dimensional vector with components
$p_m=\rho_{m,m}$, we have the Pauli equation \cite{Pauli28}
\begin{align}
  \label{Pauli}%
  \D \bm{p}(t)/\D t = -\bm{A}\ \bm{p}(t),
\end{align}
where $\bm{A}$ is the $M \times M$ matrix with components
\begin{align}
  \label{Amn}%
  &A_{m,n} = B_m \delta_{m,n} - |\ell_{m,n}|^2,
  \\
  \label{Bm}%
  &B_m = \sum_j |\ell_{j,m}|^2.
\end{align}
Equation~(\ref{Pauli}) is a master equation for the populations $p_m$,
$\D p_{m}/\D t = \sum_{n } \left[ p_{n } W_{n \to m} -p_{m} W_{m \to
    n} \right]$, with probability rates $W_{m \to
  n}=|\ell_{n,m}|^2$. We conclude that $|\ell_{n,m}|^2$ represents the
probability rate of the $|m\rangle\to|n\rangle$ transition's occurring
as a consequence of the interaction with the reservoir.  In the
weak-coupling limit, these rates can be calculated using the
time-dependent perturbation theory. In this way, the matrix elements
$C_{m,n}$ of Eq.~(\ref{ellemn}) can be fully determined.

The $M(M-1)$ off-diagonal elements of $\bm{\rho}$ evolve independently
of each other as
\begin{align}
  \label{decoherences}%
  \D \rho_{m,n}(t)/\D{t} = -\mu_{m,n} \rho_{m,n}(t), \qquad m \neq n,
\end{align}
i.e., they relax to 0 at rates
\begin{align}
  \label{mumn}%
  \mu_{m,n} = \frac{\I}{\hbar} (E'_{m}-E'_{n}) + \frac{1}{2} \left(
    B_{m} + B_{n} \right).
\end{align}

\subsection{Thermalization time} The characteristic relaxation time by
which $\bm{\rho}(t)$ approaches $\bm{\rho}^{(\mathrm{eq})}$---the
thermalization time, for brevity---depends on both Eq.~(\ref{Pauli})
and Eq.~(\ref{decoherences}).

The evolution described by Eq.~(\ref{Pauli}) is determined by the
spectrum of the matrix $\bm{A}$.  One can prove~\cite{therm} some
general properties of $\bm{A}$, independently of the particular values
assumed by the matrix elements $C_{m,n}$.  First of all, $\bm{A}$ is
diagonalizable and has $M$ real eigenvalues, possibly degenerate. It
certainly has a zero eigenvalue corresponding to the imposed
stationary solution, by virtue of $\sum_{m} A_{m,n}=0$, a
manifestation of the fact that $\mathrm{tr} \bm{\rho}(t)$ is constant.
Under fair conditions of the elements $C_{m,n}$, which are expected to
hold true for a nondegenerate $\bm{H}$, this eigenvalue is
nondegenerate, i.e., property (p1) is satisfied, whereas all the other
eigenvalues are positive.  In conclusion, the $M$ eigenvalues of
$\bm{A}$ can be ordered as $0=\mu_1(\bm{A}) < \mu_2(\bm{A})\leq \ldots
\leq \mu_M(\bm{A})$, and the relaxation time characterizing
Eq.~(\ref{Pauli}) is $1/\mu_2(\bm{A})$.

Concerning the relaxation times of the off-diagonal elements
$\rho_{m,n}$, $m\neq n$, as stated by Eq.~(\ref{mumn}), these are
trivially given by $1/\RE \mu_{m,n}$.  The largest among these times
is $2/\min_{m\neq n}\{B_m+B_n\}=2/(B_{(1)}+B_{(2)})$, where $B_{(k)}$
indicates the $k$th smallest value among the set
$\{B_1,B_2,\dots,B_M\}$.

We conclude that the thermalization time $\tau$ of our system can be
defined as
\begin{subequations}
  \label{tau_all}%
  \begin{align}
    \label{tau}%
    &\tau=\max\left\{\tau^{(P)},\tau^{(Q)}\right\},
    &&\quad \mbox{thermalization time};\\
    \label{tauP}%
    &\tau^{(P)}=1/\mu_2(\bm{A}),
    &&\quad \mbox{dissipation time};\\
    \label{tauQ}%
    &\tau^{(Q)}=2/\left( B_{(1)}+B_{(2)} \right), &&\quad
    \mbox{decoherence time}.
  \end{align}
\end{subequations}
The natural interpretation is that $\tau^{(P)}$ represents a
characteristic time by which the system exchanges energy with the
environment, whereas $\tau^{(Q)}$ represents a characteristic time by
which the system loses quantum coherence due to the interaction with
the environment.

\section{RESULTS AND DISCUSSION}
\subsection{Coupling a single system to blackbody radiation}
We can apply the above general scheme to the case in which the
environment is a blackbody radiation and the system consists of $K$
qubits (spin $1/2$). The system-environment interaction is mediated by
emission and absorption of photons via the (squared) dipole matrix
elements $D_{m,n}$. By using $|\ell_{n,m}|^2=W_{m \to n}$, we have
\cite{Davydov}
\begin{align}
  \label{Cmn}%
  C_{m ,n} = D_{m,n} \frac{|E_{m}-E_{n}|^3}
  {2\sinh\left(\frac{\beta}{2}|E_{m}-E_{n}| \right)},
\end{align}
with
\begin{align}%
  \label{Dmn.coherent}
  D_{m,n} &= \gamma\sum_{h=x,y,z} \left|\bra{m} \sum_{i=1}^{K}
    \sigma_i^{h} \ket{n}\right|^2, \quad \gamma=\frac{4\mu^2}{3\hbar^4
    c^3}.
\end{align}
Note that $C_{m,m}=0$.  Equation~(\ref{Dmn.coherent}) is obtained
supposing that the size of the system is small with respect to the
radiation length $2\pi \hbar c/|E_m-E_n|$, so that a fully coherent
interaction between spins and radiation takes place. More general
expressions can be adopted for partially or fully incoherent
interactions~\cite{goldenrule}.

\subsection{Coupling $N$ distinguishable noninteracting systems to
  blackbody radiation}
We now focus on the thermalization of an ensemble of $N$ systems,
mutually noninteracting and distinguishable. For simplicity, we
consider all equal systems, the results being easily generalized to
mixtures.  Let us examine in detail the case $N=2$.  The Hamiltonian
of the ensemble is $\bm{H}^{(2)} = \bm{H}\otimes \bm{I} +
\bm{I}\otimes \bm{H}$, where $\bm{H}$ is the $M\times M$ Hamiltonian
matrix of the single system having a nondegenerate spectrum and
$\bm{I}$ the $M\times M$ identity matrix.  We have $\bm{H}^{(2)}
\ket{m,n} = E_{m,n} \ket{m,n}$ with $E_{m,n}=E_m+E_n$ and
$\ket{m,n}=\ket{m}\otimes\ket{n}$. Note that the eigenvalues of
$\bm{H}^{(2)}$ are degenerate, even if those of $\bm{H}$ are not.  It
follows that we could choose orthonormal systems of eigenvectors
different from $\{\ket{m,n}\}$.  For instance, we have
$E_{m,n}=E_{n,m}$ and the corresponding subspace spanned by
$\ket{m,n}$ and $\ket{n,m}$, could be equivalently spanned by the Bell
states $(\ket{m,n} \pm \ket{n,m})/\sqrt{2}$.  However, this new
eigenbasis, as well as any other eigenbasis with the exception of the
product states $\{\ket{m,n}\}$, would introduce an effective
system-system correlation leading to a violation of property (p3).
This will be explicit in the decoupling of Eq.~(\ref{C_decoupling}),
below, which can take place \textit{only} for the basis
$\{\ket{m,n}\}$.  We conclude that $\{\ket{m,n}\}$ is the unique
eigenbasis of $\bm{H}^{(2)}$ in which (p3) can be satisfied, and the
LBA applied as before.

% Note that this special eigenbasis can be selected, at least
% conceptually, also in the following way.  If the two noninteracting
% systems are different, their ensemble has, in general, a
% nondegenerate energy spectrum corresponding to a unique set of
% eigenvectors in the form of product states
% $\ket{\tilde{m},\tilde{n}}=\ket{\tilde{m}}\otimes\ket{\tilde{n}}$,
% where $\ket{\tilde{m}}$ and $\ket{\tilde{n}}$ are eigenstates of the
% two different systems.  In the limit in which the differences among
% the systems fade away, $\{\ket{\tilde{m},\tilde{n}} \}\to
% \{\ket{m,n}\}$, i.e., we recover the product states of the ensemble
% of distinguishable equal systems.

In parallel to Eq.~(\ref{Lpq}), we now choose jump operators
\begin{align}
  \label{Lmnpq}%
  \bm{L}_\alpha \to \bm{L}_{m,n;p,q} =\ell_{m,n;p,q}|m,n\rangle\langle
  p,q|.
\end{align}
On imposing that the reduced density matrix operator of the ensemble
is the Gibbs SS $e^{-\beta \bm{H}^{(2)}}/\mathrm{tr}(e^{-\beta
  \bm{H}^{(2)}})$, everything follows as in the case of a single
system, just doubling the eigenstate indices.  We have
$|\ell_{m,n;p,q}|^2 =
C_{m,n;p,q}e^{-\frac{\beta}{2}(E_{m,n}-E_{p,q})}$ with
$C_{m,n;p,q}=C_{p,q;m,n}\geq 0$.  The diagonal elements
$\rho_{m,n;m,n}$ evolve according to a Pauli equation analogous to
Eq.~(\ref{Pauli}) with a matrix $\bm{A}^{(2)}$ having components
\begin{align}
  \label{Amnpq}%
  &A^{(2)}_{m,n;p,q} = B_{m,n} \delta_{m,p}\delta_{n,q} -
  |\ell_{m,n;p,q}|^2,
  \\
  \label{Bmn}%
  &B_{m,n} = \sum_{j,k} |\ell_{j,k;m,n}|^2.
\end{align}
The off-diagonal elements $\rho_{m,n;p,q}$, $m\neq p$ or $n\neq q$,
evolve according to equations analogous to Eq.~(\ref{decoherences}) at
rates
\begin{align}
  \label{mumnpq}%
  \mu_{m,n;p,q} = \frac{\I}{\hbar} (E'_{m,n}-E'_{p,q}) + \frac{1}{2}
  \left( B_{m,n} + B_{p,q} \right).
\end{align}
In the case of $N=2$ systems, each consisting of $K$ qubits coupled to
a blackbody radiation, the matrix elements $C_{m,n;p,q}$ are given by
an expression analogous to Eq.~(\ref{Cmn}) with $D_{m,n;p,q} =
\gamma\sum_{h=x,y,z} ( |\bra{m,n} \sum_{i=1}^{K} \sigma_i^{h}
\ket{p,q}|^2+ |\bra{m,n} \sum_{i=K+1}^{2K} \sigma_i^{h}
\ket{p,q}|^2)$.  It can be shown that this decomposition holds
actually in a generic eigenbasis of $\bm{H}$ if we assume that the
spin-radiation interaction is coherent within each system but the
systems are at a distance larger than the radiation length
\cite{goldenrule}.

In the basis $\{\ket{m,n}\}$, the dipole elements $D_{m,n;p,q}$
(independently of their form: coherent, incoherent or mixed) admit a
system-index decoupling and we have
\begin{align}
  \label{C_decoupling}%
  C_{m,n;p,q} = C_{m,p} \delta_{n,q} + C_{n,q} \delta_{m,p},
\end{align}
where $C_{m,n}$ are the single-system nonnegative symmetric matrix
elements, (\ref{Cmn}). Equation~(\ref{C_decoupling}) also implies that
$B_{m,n} = B_m + B_n$.  It follows that $\bm{A}^{(2)} = \bm{A}\otimes
\bm{I} + \bm{I}\otimes \bm{A}$.  It is easy to check that no
eigenbasis other than $\{\ket{m,n}\}$ permits this reduction. For
example, in the case of the Bell states, in place of
Eq.~(\ref{C_decoupling}) we would have $C_{m,n;p,q} = (C_{m,p}
\delta_{n,q} + C_{n,q} \delta_{m,p}+C_{m,q} \delta_{n,p} + C_{n,p}
\delta_{m,q})/2$.  Whereas Eq.~(\ref{C_decoupling}) corresponds to
one-body jump operators $\bm{L}_{m,n;p,q}$, the latter expression
corresponds to a mixture of two one-body jump operators where
$\bm{A}^{(2)} \neq \bm{A}\otimes \bm{I} + \bm{I}\otimes \bm{A}$.

The matrix $\bm{A}^{(2)} = \bm{A}\otimes \bm{I} + \bm{I}\otimes
\bm{A}$ has $M^2$ eigenvalues related to the $M$ eigenvalues of the
single-system matrix $\bm{A}$ by
\begin{align}
  \mu_{j,k}(\bm{A}^{(2)}) = \mu_{j}(\bm{A}) + \mu_{k}(\bm{A}), \quad
  j,k=1,2,\dots,M.
\end{align}
Since $\mu_{1}(\bm{A})=0$ is nondegenerate, $\bm{A}^{(2)}$ has a
unique zero eigenvalue. All the other eigenvalues of $\bm{A}^{(2)}$
are positive, and the smallest one of them coincides with
$\mu_{2}(\bm{A})$.  This smallest non zero eigenvalue of
$\bm{A}^{(2)}$ is $2d$-fold degenerate, if $d$ is the degeneracy of
$\mu_2(\bm{A})$.

From Eq.~(\ref{mumnpq}) we get
\begin{align}
  \min_{m\neq p\ \mathrm{or}\ n\neq q} \RE \mu_{m,n;p,q} = 3 B_{(1)} +
  B_{(2)}.
\end{align}

The above analysis is straightforwardly extended to an ensemble of $N$
all equal systems with Hamiltonian $\bm{H}^{(N)} = (\bm{H} \otimes
\bm{I} \otimes \dots \otimes \bm{I}) + (\bm{I} \otimes \bm{H} \otimes
\dots \otimes \bm{I}) + \dots +(\bm{I} \otimes \bm{I} \otimes \dots
\otimes \bm{H})$.  The result is: $\tau^{(P)}=1/\mu_2(\bm{A})$
independent of $N$, and $\tau^{(Q)} = 2/((2N-1)B_{(1)}+B_{(2)})$,
decreasing as $1/N$. For a sufficiently large $N$ and finite
temperature, we have $\tau=\max\{\tau^{(P)},\tau^{(Q}\}=\tau^{(P)}$,
independent of $N$.

\subsection{Free spins in a magnetic field}  As an example, consider a
system of $N$ independent spins located at different fixed positions,
in the presence of a uniform magnetic field of strength $\Gamma$
oriented along the $x$ axis. The Hamiltonian of the single spin is
$\bm{H} = -\Gamma \sigma^{x}$ and has a nondegenerate spectrum,
namely, $E_1=-\Gamma$ and $E_2=\Gamma$ with eigenvectors
$\ket{1}=\ket{+}$ and $\ket{2}=\ket{-}$, where $\sigma^x \ket{\pm}=\pm
\ket{\pm}$.  We have $D_{1,2}=D_{2,1}=2\gamma$ and
$C_{1,2}=C_{2,1}=2\gamma
(2\Gamma)^3/(e^{\beta\Gamma}-e^{-\beta\Gamma})$, which lead to
\begin{align}
  \bm{A} = \frac{2\gamma
    (2\Gamma)^3}{e^{\beta\Gamma}-e^{-\beta\Gamma}} \left(
    \begin{array}{cc}
      e^{-\beta\Gamma} & -e^{\beta\Gamma}
      \\
      -e^{-\beta\Gamma} & e^{\beta\Gamma}
    \end{array}
  \right).
\end{align}
The two eigenvalues of $\bm{A}$, both nondegenerate, are
$\mu_1(\bm{A})=0$ and $\mu_2(\bm{A})=(2\gamma
(2\Gamma)^3)/\tanh(\beta\Gamma)$.  Since the $N$ spins are
distinguishable by virtue of their different fixed position, we
conclude that
\begin{align}
  \label{tauP_FS}%
  \tau^{(P)} &= \frac{\tanh(\beta\Gamma)}{2\gamma(2\Gamma)^3},
  \\
  \label{tauQ_FS}%
  \tau^{(Q)} &= \frac{\sinh(\beta\Gamma)}{\gamma(2\Gamma)^3
    \left(\sinh(\beta\Gamma) + Ne^{-\beta\Gamma}\right)}.
\end{align}

As a further example of independent systems with different spectra, in
Appendix~\ref{QOME-appendix} we provide the expressions of
$\tau^{(P)}$ and $\tau^{(Q)}$ for the above ensemble of $N$ spins in a
spatially periodic magnetic field.  In this case, $\tau^{(P)}$ becomes
independent of $N$ only for $N$ large.

\subsection{Coupling $N$ indistinguishable noninteracting systems to
  blackbody radiation}
In this case, the laws of quantum mechanics select either totally
symmetric or totally anti-symmetric eigenvectors, if the systems are
bosonic or fermionic, respectively. As a consequence, the degeneracy
of the spectrum of $\bm{H}^{(N)}$ is greatly reduced even if not
completely eliminated.  Consider, for example, a system (bosonic or
fermionic) whose Hamiltonian $\bm{H}$ has four levels
$E_1<E_2<E_3<E_4$ such that $E_1+E_4=E_2+E_3$.  The Hamiltonian
$\bm{H}^{(2)}$ of the ensemble has symmetrized or antisymmetrized
eigenvectors which are twofold degenerate.  However, the origin of
this degeneracy is different from that of the unavoidable degeneracies
which appear for distinguishable systems.  It is due to the occurrence
of particular relations among the levels of $\bm{H}$, which have to be
regarded as rather uncommon. Typically, the basis of $\bm{H}^{(N)}$ is
unique and the LBA well defined.  As observed, however, this unique
eigenbasis is not made by product states of single-system eigenstates,
and, from what we have learned in the case of distinguishable systems,
only (p1) and (p2) can be satisfied, whereas $\tau$ can be independent
of $N$ only in the limit of low densities.  Yet, property (p4) holds
unchanged~\cite{therm}, and we expect that also (p5) and (p6) still
apply.

\textit{Pathologies of the microscopic derivation.}  In the Lindblad
equation resulting from the microscopic derivation, for an ensemble of
$N$ noninteracting systems described by the Hamiltonian
$\bm{H}^{(N)}$, instead of Eq. (\ref{Lpq}) one has (see Eqs. (3.120)
and (3.143) in Ref.~\cite{Petruccione})
\begin{align}
  \label{LpqQOME}%
  \bm{L}_\alpha \to \bm{A}(\omega) = \hspace{-2em}
  \sum_{\overset{m^{(N)},n^{(N)}:}{E_{n^{(N)}}-E_{m^{(N)}}=\hbar\omega}}
  \hspace{-2em} A_{m^{(N)},n^{(N)}} \ket{m^{(N)}}\bra{n^{(N)}},
\end{align}
where $\bm{H}^{(N)} \ket{m^{(N)}} = E_{m^{(N)}} \ket{m^{(N)}}$.  The
symbol $\bm{A}(\omega)$ is borrowed from~\cite{Petruccione} and should
not be confused with the matrix $\bm{A}$ used before in our Pauli
equation.  Note that, even in the presence of level degeneracies, the
basis $\{\ket{m^{(N)}}\}$ used in Eq.~(\ref{LpqQOME}) is arbitrary. In
fact, the operator $\bm{A}(\omega)$ is invariant with respect to a
change of the eigenvectors of $\bm{H}^{(N)}$.  For simplicity, let us
focus on $N=2$ noninteracting, equal but distinguishable, systems,
each having a nondegenerate spectrum with two levels, $E_1<E_2$.  As
the eigenbasis to be used in Eq.~(\ref{LpqQOME}), let us choose the
product eigenstates, $|m^{(2)}\rangle =|m,n\rangle=|m\rangle\otimes
|n\rangle$, with $m,n\in\{1,2\}$.  There are only two possible
positive values of $\omega$ (and two corresponding negative values for
which identical considerations hold): $\hbar\omega_1=E_2-E_1$ and
$\hbar\omega_2=2(E_2-E_1)$.  Whereas for the latter we have
$\bm{A}(\omega_2)=A_{1,1;2,2} |1,1\rangle\langle 2,2|$, for the former
we find $\bm{A}(\omega_1)= A_{1,1;1,2} |1,1\rangle\langle 1,2|+
A_{1,1;2,1} |1,1\rangle\langle 2,1| +A_{1,2;2,2} |1,2\rangle\langle
2,2|+ A_{2,1;2,2} |2,1\rangle\langle 2,2|$.  We see that each Lindblad
operator $\bm{A}(\omega)$ can be the sum of more dyadic operators and
this happens in correspondence with level degeneracies.  By a similar
example with three energy levels, it is easy to see that also gap
degeneracies lead to extra dyadic operators. In other words,
Eq.~(\ref{LpqQOME}) introduces correlations among the single-system
eigenstates which can lead to violations of properties
(p1)--(p3). Ultimately, these correlations are an artifact due to the
approximations used in the microscopic derivation.

In the most general case of $N$ distinguishable systems, possibly also
different, the Lindblad equation obtained in our LBA, coincides with
the Lindblad equation of the microscopic derivation only when both the
conditions $\delta_{E_m,E_n,}=\delta_{m,n}$ and
$\delta_{E_m-E_n,E_p-E_q}=\delta_{m,p}\delta_{n,q}$ are met.  These
two conditions ensure that there are no level or gap degeneracies.
Following Ref.~\cite{PhysRevE.94.022150}, these can be interpreted as
the conditions required for the isolated many-body system represented
by the Hamiltonian $\bm{H}^{(N)}$ to be ergodic and mixing,
respectively.  These conditions are never satisfied in the case of
$N>1$ equal systems.  In Appendix~\ref{QOME-appendix} we provide a
detailed comparison, also with numerical examples, between the quantum
optical master equation, which is the Lindblad equation obtained in
the microscopic derivation when the environment is a blackbody
radiation, and the corresponding Lindblad equation obtained within our
approach.

In the case of indistinguishable noninteracting systems, even if we
exclude accidental degeneracies, in Eq.~(\ref{LpqQOME}) we still have
extra terms as soon as there are gap degeneracies, and it is easy to
see that, for bosons, the latter always occur.  An extended analysis
of the case of indistinguishable systems will be reported elsewhere.

\section{CONCLUSION}
In conclusion, we have first considered the thermalization of
arbitrary systems with a nondegenerate spectrum and in contact with a
thermal reservoir. The analysis is done by using a Lindblad-based
approach in which the Lindblad operators, initially chosen as a
completely general basis set, are next fixed by requiring the
existence of a Gibbs stationary state. This approach allows us to
clearly identify two characteristic times $\tau^{(P)}$ and
$\tau^{(Q)}$, namely, the dissipation and decoherence times, the
largest of which determines the thermalization time.  We have then
applied this procedure to ensembles of $N$ mutually noninteracting
subsystems coupled to a blackbody radiation and found explicit
formulas for $\tau^{(P)}$ and $\tau^{(Q)}$.  For ensembles of
distinguishable subsystems, the approach must be equipped with the
phenomenological condition that the thermalization time should not
depend on the ensemble size $N$, which leads to complete determination
of the Lindblad operators despite the exchange degeneracies.  The
thermalization of $N$ free spins in a magnetic field, uniform or not,
coupled to a blackbody radiation, has been considered in detail.  For
this system, techniques for solving the Lindblad equation like those
in \cite{Prosen_2008} and \cite{Prosen_2010} are
inapplicable~\footnote{ One can see this by using a Jordan-Wigner
  transformation.  At most two components of each spin are mapped into
  expressions linear into creation and annihilation fermionic
  operators, while the third spin component will be expressed as a
  product of these creation and annihilation operators.  It then
  follows that the Lindblad operators representing the dipole
  interactions of the electromagnetic field with the three spin
  components is not linear in the creation and annihilation fermionic
  operators as required in~\cite{Prosen_2008}}, yet it is
straightforwardly and consistently worked out by our approach. An
analysis of the same system within the quantum optical master equation
reveals serious physical inconsistencies.

\appendix

\section{Lindblad-based approach to thermalization}
\label{LBA-appendix}
Let us denote $\bra{m}\bm{\rho}\ket{n}=\rho_{m,n}$ and
$\bra{m}\bm{H}'\ket{n}=H'_{m,n}$, with $\bm{H} \ket{m} = E_m \ket{m}$.
Equation~(\ref{GEN_LIN}), with the choice of (\ref{Lpq}), becomes
\begin{widetext}
\begin{align}
  \label{LIN_Lpq}%
  \frac{\D \rho_{m,n}}{\D t} = &-\frac{\I}{\hbar} \sum_{k } \left(
    H'_{m ,k }\rho_{k ,n } - \rho_{m ,k } H'_{k ,n } \right) +\sum_{k
  } \left( \left| \ell_{m ,k } \right|^2 \rho_{k ,k } \delta_{m,n}
    -\frac{1}{2} \left| \ell_{k ,m } \right|^2 \rho_{m ,n }
    -\frac{1}{2} \left| \ell_{k ,n } \right|^2 \rho_{m,n} \right).
\end{align}
On imposing that $\bm{\rho}^{(\mathrm{eq})} = e^{-\beta
  \bm{H}}/\mathrm{tr}(e^{-\beta \bm{H}})$ is a stationary solution of
Eq.~(\ref{LIN_Lpq}), we get
\begin{align}
  \label{LIN_Lpq_eq}%
  0 =& -\frac{\I}{\hbar} \left( H'_{m,n} e^{-\beta E_{n }} - e^{-\beta
      E_{m}} H'_{m,n} \right) + \sum_{k } \left( \left| \ell_{m ,k }
    \right|^2 e^{-\beta E_{k }} - \left| \ell_{k ,m } \right|^2
    e^{-\beta E_{m}} \right) \delta_{m,n}.
\end{align}
\end{widetext}

For $m \neq n$, since $E_m \neq E_n$, Eq.~(\ref{LIN_Lpq_eq}) implies
that $H'_{m,n}=0$.  We thus infer that $H'_{m,n} = E'_{m} \delta_{m,n}$,
with $E'_m$ real as ${\bm{H}'}^\dag=\bm{H}'$. In other words,
$\bm{H}'$ is diagonal in the basis of the eigenvectors of $\bm{H}$.
We still ignore the eigenvalues $E'_{m }$ but this is not relevant for
determining the relaxation time to equilibrium.

For $m=n$, the purely imaginary term on the right-hand side of
Eq.~(\ref{LIN_Lpq_eq}) vanishes and we deduce that the coefficients
$\ell_{m,n}$ must satisfy the balance condition
\begin{align}
  \label{LINEQ}%
  \sum_{k } \left| \ell_{m ,k } \right|^2 e^{-\beta E_{k }} = \sum_{k
  } \left| \ell_{k ,m } \right|^2 e^{-\beta E_{m}}.
\end{align}
The most general detailed-balance solution of Eq.~(\ref{LINEQ}) can be
written as
\begin{align}
  \label{LIN2}%
  |\ell_{m,n}|^2 = C_{m,n} e^{-\frac{\beta}{2} (E_{m}-E_{n })},
\end{align}
provided $C_{m,n}=C_{n,m}\geq 0$.

We conclude that the elements $\rho_{m,n}$ of the reduced density
matrix of the system evolve according to the following system of
equations
\begin{widetext}
\begin{align}
  \label{LIN3G}%
  \frac{\D \rho_{m,n}}{\D{t}} = \left[ -\frac{\I}{\hbar} (E'_{m}-E'_{n
    }) -\frac{1}{2} \sum_k \left( |\ell_{k,m}|^2 + |\ell_{k,n}|^2
    \right) \right] \rho_{m,n} + \sum_{k } |\ell_{m,k}|^2 \rho_{k ,k }
  \delta_{m,n}.
\end{align}
It follows that the evolution of the diagonal components $\rho_{m,m}$
is decoupled from that of the off-diagonal elements, and the latter
are also decoupled fromeach other:
\begin{align}
  \label{LIN3G.1}%
  \frac{\D \rho_{m,m}}{\D{t}} = - \left(\sum_k |\ell_{k,m}|^2 \right)
  \rho_{m,m} + \sum_{k} |\ell_{m,k}|^2 \rho_{k,k},
\end{align}
\begin{align}
  \label{LIN3G.2}%
  \frac{\D \rho_{m,n}}{\D{t}} = \left[ -\frac{\I}{\hbar} (E'_m-E'_n)
    -\frac{1}{2} \sum_k \left( |\ell_{k,m}|^2 + |\ell_{k,n}|^2 \right)
  \right] \rho_{m,n}, \qquad m \neq n.
\end{align}
\end{widetext}

\section{Coupling with blackbody radiation: Lindblad-based approach vs
  microscopic derivation}
\label{QOME-appendix}
Equation~(\ref{LIN3G}), customized with $|\ell_{m,n}|^2 = C_{m,n}
e^{-\frac{\beta}{2} (E_{m}-E_{n })}$ and the coefficients $C_{m,n}$
given in Eq.~(\ref{Cmn}), is the optimal Lindblad equation obtained
within our LBA in the case of coupling with a blackbody radiation.
Explicitly, this equation reads
\begin{align}
  \label{LBA}%
  \frac{\D\rho_{m,n}}{\D{t}} =& -\frac{\I}{\hbar} (E'_m -E'_n)
  \rho_{m,n} \nonumber \\
  &-\frac{1}{2} \sum_{k} \left( D_{k,m} \widetilde{W}_{k,m} + D_{k,n}
    \widetilde{W}_{k,n} \right) \rho_{m,n} \nonumber \\ &+ \sum_{k }
  D_{m,k} \widetilde{W}_{m,k} \rho_{k,k} \delta_{m,n},
\end{align}
where
\begin{align}
  \widetilde{W}_{m,k}= |E_{m}-E_{k }|^3 \frac{e^{-\frac{\beta}{2}
      (E_{m}-E_{k})}}
  {e^{\frac{\beta}{2}|E_{m}-E_{k}|}-e^{-\frac{\beta}{2}|E_{m}-E_{k}|}}.
\end{align}
For a system of $N$ qubits, the (squared) dipole matrix elements
$D_{m,n}$ evaluated in the fully coherent limit are
\begin{align}
  \label{Dmn.coherent.SM}%
  D_{m,n} = \frac{4\mu^2}{3\hbar^4 c^3} \sum_{h=x,y,z} \left|\bra{m}
    \sum_{i=1}^{N} \sigma_i^{h} \ket{n}\right|^2.
\end{align}
Expressions more general than Eq.~(\ref{Dmn.coherent.SM}) can be
adopted, depending on the size of the system; see
Ref.~\cite{goldenrule}.

In the microscopic derivation, the coupling of a system with a
blackbody radiation is described by the celebrated quantum optical
master equation~\cite{Petruccione}; see also~\cite{Schaller} for an
alternative derivation. This equation presents some differences with
respect to the LBA, Eq.~(\ref{LBA}).  Consider Eq.~(3.206) in
Ref.~\cite{Petruccione}, which provides a general quantum optical
master equation in the Lindblad operatorial form and in the
interaction picture.  Switching to the Schr\"odinger picture and
taking the matrix element of the corresponding operators between
eigenstates $\ket{m}$ and $\ket{n}$ of the same $\bm{H}$ considered
above, we get
\begin{widetext}
  \begin{align}
    \label{QOME}%
    \frac{\D\rho_{m,n}}{\D{t}} =& -\frac{\I}{\hbar} (E_m -E_n)
    \rho_{m,n} \nonumber\\ &-\frac{\I}{\hbar} \sum_{k,j} \left[ \left(
        \hbar S(E_m-E_k) \sum_{h=x,y,z} d^{(h)}_{k,j}
        \overline{d^{(h)}_{k,m}}\right) \rho_{j,n} \delta_{E_j,E_m} -
      \left( \hbar S(E_n-E_k) \sum_{h=x,y,z} d^{(h)}_{k,n}
        \overline{d^{(h)}_{k,j}}\right) \rho_{m,j} \delta_{E_j,E_n}
    \right] \nonumber\\ &- \frac{1}{2} \sum_{k,j} \left[
      \left(\frac{4\mu^2}{3\hbar^4 c^3} \sum_{h=x,y,z} d^{(h)}_{k,j}
        \overline{d^{(h)}_{k,m}}\right) \widetilde{W}_{k,m} \rho_{j,n}
      \delta_{E_j,E_m} + \left(\frac{4\mu^2}{3\hbar^4 c^3}
        \sum_{h=x,y,z} d^{(h)}_{k,n} \overline{d^{(h)}_{k,j}}\right)
      \widetilde{W}_{k,n} \rho_{m,j} \delta_{E_j,E_n} \right]
    \nonumber\\ &+ \sum_{k,j} \left(\frac{4\mu^2}{3\hbar^4 c^3}
      \sum_{h=x,y,z} d^{(h)}_{m,k} \overline{d^{(h)}_{n,j}}\right)
    \widetilde{W}_{m,k}\rho_{k,j} \delta_{E_k-E_m,E_j-E_n},
  \end{align}
  where
  \begin{align}
    \label{SE}%
    S(E)= \frac{2\mu^2|E|^3}{3\pi\hbar^4c^3} \mathrm{PV}
    \int_{0}^{\infty} \left( \frac{u^3}{1-e^{-\beta Eu}}~\frac{1}{1-u}
      + \frac{u^3}{e^{\beta Eu}-1}~\frac{1}{1+u}\right) \D{u}
  \end{align}
\end{widetext}
and
\begin{align}
  d^{(h)}_{m,n} = \bra{m} \sum_{i=1}^{N}\sigma_i^{h} \ket{n} .
\end{align}

First, we observe that in the quantum optical master equation the
contribution from the Lamb shift Hamiltonian is given explicitly by
the second line of Eq.~(\ref{QOME}).  This term, which in general is
not diagonal in the basis $\{|m\rangle \}$ due to the presence of the
two Kronecker deltas, $\delta_{E_j,E_m} \neq \delta_{j,m}$ and
$\delta_{E_j,E_n} \neq \delta_{j,n}$, is given in terms of the Cauchy
principal value of the integral (\ref{SE}).  The integral presents an
ultraviolet ($u\to\infty$) divergence, presumably introduced by the
approximations used to derive the quantum optical master equation.  To
meaningfully compare Eq.~(\ref{QOME}) with Eq.~(\ref{LBA}), we neglect
the Lamb shift correction to the levels of $\bm{H}$.  This amounts to
disregarding the entire second line of Eq.~(\ref{QOME}) and putting
$E_m'=E_m$ in Eq.~(\ref{LBA}).  As far as Eq.~(\ref{LBA}) is
concerned, we have proven that such a change does not alter the
relaxation time to equilibrium, since neither $\tau^{(P)}$ nor
$\tau^{(Q)}$ depends on $E'_m$.

If the eigenvalues of $\bm{H}$ satisfy
\begin{align}
  \label{ergodicity.Supp}%
  \delta_{E_j,E_m} = \delta_{j,m}
\end{align}
and
\begin{align}
  \label{mixing.Supp}%
  \delta_{E_k-E_m,E_j-E_n} = \delta_{k,j}\delta_{m,n},
\end{align}
Eq.~(\ref{QOME}) reduces to Eq.~(\ref{LBA}) with the coefficients
$D_{m,n}$ evaluated in the fully coherent limit given by
Eq.~(\ref{Dmn.coherent.SM}) To the best of our knowledge,
Eq.~(\ref{QOME}) is known only in a formulation compatible with a
fully coherent dipole interaction, i.e., a formulation valid when the
$N$ qubits are at reciprocal distances shorter than the radiation
lengths $2\pi\hbar c/|E_m-E_n|$.

\begin{figure}[h]
  \begin{center}
    \includegraphics[width=0.99\columnwidth,clip]{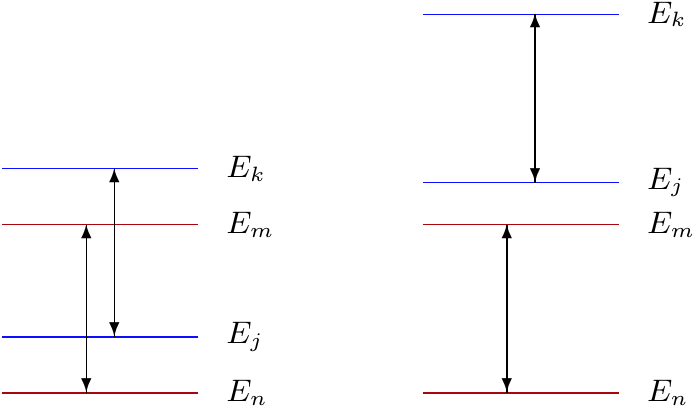}
    \caption{(Color online) Two possible arrangements of the four
      energy levels $E_m$, $E_n$, $E_j$, and $E_k=E_m-E_n+E_j$,
      providing terms of Eq.~(\ref{QOME}) extraneous to
      Eq.~(\ref{LBA}). Here, we have assumed a non-degenerate spectrum
      with gap degeneracies (vertical double-arrows), and $m \neq n$.}
    \label{levels.QOME}%
  \end{center}
\end{figure}

If the only condition (\ref{ergodicity.Supp}) is satisfied,
Eq.~(\ref{QOME}) splits into a Pauli equation for the $M$ diagonal
terms $\rho_{m,m}$, which is identical to the Pauli equation obtained
in the LBA, and a set of equations for the off-diagonal terms
$\rho_{m,n}$, $m \neq n$.  In fact, in this case, for $m=n$ we have
$\delta_{E_k-E_m,E_j-E_n}=\delta_{E_k,E_j}=\delta_{k,j}$ and the last
line of Eq.~(\ref{QOME}) coincides with the last term of
Eq.~(\ref{LBA}), whereas for $m \neq n$, i.e., $E_m \neq E_n$,
$\delta_{E_k-E_m,E_j-E_n}$ implies that $E_k \neq E_j$ and, therefore,
$k \neq j$.  However, unlike the equations obtained in the LBA, the
$M(M-1)$ equations for the off-diagonal elements of $\bm{\rho}$ are
mutually coupled.  In Fig.~\ref{levels.QOME} we show examples of
possible arrangements of four energy levels which satisfy
Eq.~(\ref{ergodicity.Supp}) but violate Eq.~(\ref{mixing.Supp}).

If neither of conditions~(\ref{ergodicity.Supp}) and
(\ref{mixing.Supp}) is satisfied, Eq.~(\ref{QOME}) is a set of $M^2$
linear coupled equations for the $M^2$ elements $\rho_{m,n}$. The
thermalization times can be found by diagonalizing the Liouvillian
\begin{widetext}
  \begin{align}
    \label{QOME.Liouvillian}%
    \mathcal{L}_{m,n;k,j} =& -\frac{\I}{\hbar} (E_m -E_n) \delta_{m,k}
    \delta_{n,j} \nonumber\\ &- \frac{1}{2} \sum_q
    \left(\frac{4\mu^2}{3\hbar^4 c^3} \sum_{h=x,y,z} d^{(h)}_{q,k}
      \overline{d^{(h)}_{q,m}} \right) \widetilde{W}_{q,m}
    \delta_{E_k,E_m} \delta_{n,j} - \frac{1}{2} \sum_q
    \left(\frac{4\mu^2}{3\hbar^4 c^3} \sum_{h=x,y,z} d^{(h)}_{q,n}
      \overline{d^{(h)}_{q,j}}\right) \widetilde{W}_{q,n}
    \delta_{E_j,E_n} \delta_{m,k} \nonumber\\ &+
    \left(\frac{4\mu^2}{3\hbar^4 c^3} \sum_{h=x,y,z} d^{(h)}_{m,k}
      \overline{d^{(h)}_{n,j}}\right) \widetilde{W}_{m,k}
    \delta_{E_k-E_m,E_j-E_n}.
  \end{align}
\end{widetext}
After vectorization, the Liouvillian (\ref{QOME.Liouvillian}) is an
$M^2 \times M^2$ matrix with a zero eigenvalue, possibly degenerate,
and with complex eigenvalues with negative real parts.  The
dissipation time can be defined as $\tau^{(P)}=-1/\mu$, where $\mu$ is
a real eigenvalue, possibly degenerate, with the smallest non zero
modulus.  The decoherence time can be obtained as
$\tau^{(Q)}=-1/\RE\mu$, where $\mu$ is an eigenvalue, possibly
degenerate, having nonzero imaginary part and real part with the
smallest absolute value.  For systems with nondegenerate levels but
degenerate gaps, this definition coincides with that obtained
considering the separate equations for the diagonal and off-diagonal
elements of $\bm{\rho}$.

As an example of the differences which can emerge between the
solutions of the master equations obtained within the microscopic
approach and those obtained wihin our LBA, we have considered systems
of $N$ free spins immersed in a magnetic field, uniform or not,
described by the Hamiltonian
\begin{align}
  \label{HFSnonU}
  \bm{H} = -\sum_{i=1}^N \Gamma_i \sigma_i^x.
\end{align}
The LBA thermalization times can be found analytically, namely,
\begin{align}
  \label{tauP_FSnonU}%
  \tau^{(P)} &= \max_{i=1,\dots,N}
  \frac{\tanh(\beta\Gamma_i)}{2\gamma\left(2\Gamma_i\right)^3},
\end{align}
\begin{align}
  % \\
  \label{tauQ_FSnonU}%
  \tau^{(Q)} &= \max_{i=1,\dots,N} \left(
    \gamma\left(2\Gamma_i\right)^3
    \frac{\cosh(\beta\Gamma_i)}{\sinh(\beta\Gamma_i)} \phantom{\sum_{k
        \neq i}} \right. \nonumber \\ &\qquad\qquad+ \left.  \sum_{k
      \neq i} \gamma\left(2\Gamma_k\right)^3
    \frac{e^{-\beta\Gamma_k}}{\sinh(\beta\Gamma_k)} \right)^{-1}.
\end{align}
  
In the uniform case $\Gamma_i=\Gamma$, the spectrum of $\bm{H}$ is
degenerate and has gap degeneracies, i.e., neither of the
conditions~(\ref{ergodicity.Supp}) and (\ref{mixing.Supp}) is
satisfied.  The numerical diagonalization of the Liouvillian,
(\ref{QOME.Liouvillian}), reveals a multiplicity of zero eigenvalues
which increases with $N$, as well as, a dissipation time depending on
$N$ and a decoherence time not decreasing as $1/N$.  These features
are in conflict with properties (p1)--(p3) and the explicit results of
Eqs.~(\ref{tauP_FSnonU}) and (\ref{tauQ_FSnonU}).
  
In the nonuniform case, the differences between the LBA and the
microscopic approach are milder.  The spectrum of $\bm{H}$ is, in
general, nondegenerate but has gap degeneracies, due to a parity
symmetry.  The numerical diagonalization of the Liouvillian,
(\ref{QOME.Liouvillian}), reveals a unique zero eigenvalue, a
dissipation time equal to that obtained from our Eq.~(\ref{LBA}), but
a decoherence time still not decreasing as $1/N$, at least for the
small $N$ explored. In Table~\ref{table_FSnonU} we detail the results
obtained for a magnetic field with oscillating strength
$\Gamma_i=1+\sin((i-1)\pi/\sqrt{2})/2$.
\begin{table*}
  \caption{\label{table_FSnonU}
    Thermalization of a system of $N$ free spins immersed in a nonuniform
    magnetic field and coupled to a blackbody radiation at inverse 
    temperature $\beta=1$. The Hamiltonian of the isolated system is 
    given by Eq.~(\ref{HFSnonU}), with
    $\Gamma_i=1+\sin((i-1)\pi/\sqrt{2})/2$.
    As a function of the size $N$, we report $\tau^{(P)}$ and $\tau^{(Q)}$ 
    evaluated from the analytical (columns 2 and 3) 
    and numerical (columns 4 and 5) solution of our LBA Eq.~(\ref{LBA}) and 
    from the numerical solution of the quantum optical master equation 
    (QOME) (columns 7 and 8).
    Columns 6 and 9 show the cpu time used to execute 
    the LBA and QOME codes, respectively. 
    The LBA code corresponds, essentially, to the diagonalization 
    of a $2^N \times 2^N$ sparse matrix. With the computer memory at our 
    disposal (16 GB), we cannot execute this code for $N>13$. 
    In the QOME case, the matrix has dimensions $2^{2N} \times 2^{2N}$ and 
    we have to stop at $N=6$. Note that, as expected, $\tau^{(P)}$ tends 
    to a constant for $N\to\infty$, whereas $\tau^{(Q)}$ decreases as 
    $1/N$ for $N$ large. The latter behavior is not reproduced by the QOME.}  
  \begin{ruledtabular}
    \begin{tabular}{c cc ccc ccc}
      &\multicolumn{2}{c}{LBA}&\multicolumn{3}{c}{LBA numerical}&
      \multicolumn{3}{c}{QOME numerical}\\
      $N$&$\tau^{(P)}$&$\tau^{(Q)}$
      &$\tau^{(P)}$&$\tau^{(Q)}$&cpu time (s)
      &$\tau^{(P)}$&$\tau^{(Q)}$&cpu time (s)\\
      1& 0.04760 & 0.09520 & 0.04760 & 0.09520 & 0.002 & 
      0.04760 & 0.09520 & 0.002 \\
      2& 0.04760 & 0.07493 & 0.04760 & 0.07493 & 0.002 & 
      0.04760 & 0.09520 & 0.002 \\
      3& 0.21406 & 0.13016 & 0.21406 & 0.13016 & 0.002 & 
      0.21406 & 0.42813 & 0.006 \\
      4& 0.21406 & 0.09589 & 0.21406 & 0.09589 & 0.003 & 
      0.21406 & 0.42813 & 0.411 \\
      5& 0.21406 & 0.07560 & 0.21406 & 0.07560 & 0.006 & 
      0.21406 & 0.42813 & 84.22 \\
      6& 0.22800 & 0.06989 & 0.22800 & 0.06989 & 0.022 & 
      0.22800 & 0.45600 & 5696 \\
      7& 0.22800 & 0.05833 & 0.22800 & 0.05833 & 0.111 &  &  &  \\
      8& 0.22800 & 0.05058 & 0.22800 & 0.05058 & 0.546 &  &  &  \\
      9& 0.22800 & 0.04733 & 0.22800 & 0.04733 & 3.789 &  &  &  \\
      10& 0.22800 & 0.04172 & 0.22800 & 0.04172 & 28.25 &  &  &  \\
      11& 0.22800 & 0.03809 & 0.22800 & 0.03809 & 229.3 &  &  &  \\
      12& 0.22800 & 0.03573 & 0.22800 & 0.03573 & 1839 &  &  &  \\
      13& 0.22800 & 0.03245 & 0.22800 & 0.03245 & 20411 &  &  &  \\
      100& 0.23104 & 0.004433 &  &  &  &  &  &  \\
      1000& 0.23106 & 0.0004455 &  &  &  &  &  &  \\
      10000& 0.23106 & 0.00004457 &  &  &  &  &  &  \\
      100000& 0.23106 & 0.000004458 &  &  &  &  &  &  \\
    \end{tabular}
  \end{ruledtabular}
\end{table*}
  
\begin{acknowledgments}
  M.O. acknowledges the Coordena\c{c}\~{a}o 
de Aperfei\c{c}oamento de Pessoal de N\'ivel 
  Superior (CAPES), Brasil, for its PNPD program.
\end{acknowledgments}

% \bibliography{therm} % run pdflatex bibtex pdfltex pdflatex
% \input{this_file_name.bbl} % to be used before submission

% merlin.mbs apsrev4-1.bst 2010-07-25 4.21a (PWD, AO, DPC) hacked
% Control: key (0) Control: author (0) dotless jnrlst Control: editor
% formatted (1) identically to author Control: production of article
% title (0) allowed Control: page (1) range Control: year (0) verbatim
% Control: production of eprint (0) enabled
%

\end{document}